\documentclass{article}
\usepackage[utf8]{inputenc}

    \title{Symbolic Knowledge Structures and Intuitive Knowledge Structures}

\author{Nancy Lynch \\
	Massachusetts Institute of Technology \\
	Department of Electrical Engineering and
Computer Science \\
    \texttt{lynch@csail.mit.edu}\thanks{This work was supported by the National Science Foundation awards CCF-1810758 and CCF-2139936.}}
\date{\today}

\begin{document}

\maketitle

\begin{abstract}
This paper proposes that two distinct types of structures are present in the brain:  \emph{Symbolic Knowledge Structures (SKSs)}, used for formal symbolic reasoning, and \emph{Intuitive Knowledge Structures (IKSs)}, used for drawing informal associations.  
The paper contains ideas for modeling and analyzing these structures in an algorithmic style based on Spiking Neural Networks, following the paradigm of~\cite{LynchMP-arxiv19, SuCL-jour19, HitronLMP20, LynchMallmannTrenn21}.
The paper also contains two examples of use of these structures, involving counting through a memorized sequence, and understanding simple stylized sentences. 

\emph{Disclaimer:} The ideas presented here are preliminary and speculative, and do not (yet) comprise a complete, coherent, algorithmic theory.
That will require more thought and work.
I hope that posting this preliminary version will help the ideas to evolve into such a theory.
\end{abstract}

\section{Introduction}

This work was originally motivated by trying to understand how language is processed in the human brain.  Over time, the work has evolved so that now it focuses on two separate kinds of structures in the brain: one representing \emph{symbolic knowledge} and one representing \emph{intuitive knowledge}.
An \emph{intuitive knowledge structure} represents informally-understood concepts.  It supports construction of new concepts and relationships, as well as drawing informal associations between concepts, so that thoughts about some concepts may trigger thoughts about others.
A \emph{symbolic knowledge structure}, on the other hand, represents symbols (numbers, words, parts of speech, etc.), and supports formal, logical reasoning.

These two types of structures, symbolic and intuitive, are used in language processing.  A symbolic knowledge structure includes precise representations of linguistic elements such as words and parts of speech.  It also encodes formal structures such as parse trees, and other relationships between linguistic elements.
An intuitive knowledge structure includes imprecise representations of concepts that are denoted by words and sentences, such as pictures and simple stories.  It also encodes informally-understood relationships between concepts.

However, language processing isn't the only use for these structures.
In an example I develop in this paper, one might memorize a sequence of symbols, each of which corresponds to some elaborate concept; for example, the successive letters in the Greek alphabet might be used to name the successive variants of an aggressive virus, or successive hurricanes in a particular year.
Associated with each Greek letter is a collection of information about the virus variant or hurricane, which can be informally understood.  The Greek names would appear in a symbolic knowledge structure, whereas the information about the virus or hurricane would appear in an intuitive knowledge structure.

Yet another example might arise in planning a plumbing system, or other construction project.
The plumbing system designer knows intuitively that water should flow in a pipe at approximately a certain natural rate---knowledge that lives in an intuitive knowledge structure.
The designer must translate that intuitive requirement into precise angles at which to place the pipes; this part requires mathematical (symbolic) calculation, in a symbolic knowledge structure.

My approach in this paper is in terms of abstract models and algorithms, not in terms of actual areas of the human brain.
This is the approach my co-workers and I have taken previously, in studying such problems as Winner-Take-All selection~\cite{LynchMP-arxiv19,SuCL-jour19},
neural coding of complex data~\cite{HitronLMP20},
and recognition and learning of hierarchical concepts~\cite{LynchMallmannTrenn21}.
It would be nice to (eventually) relate these abstract structures to brain areas.
It seems plausible that the two kinds of structures considered here might reside in different areas of the brain, containing neurons with different characteristics.
For example, a symbolic knowledge structure should be implemented using neurons that are reliable, with reliable connections.  An intuitive knowledge structure could be built using less reliable neurons, with noisy connections.

This division into symbolic knowledge structures and intuitive knowledge structures should be useful in understanding the differences between the intelligence of humans and other animals.  Basically, both humans and many animals utilize intuitive knowledge and computation to understand their world and plan their activities.
Humans also make extensive use of symbolic knowledge and computation.

\section{Model for Symbolic and Intuitive Knowledge}
\label{sec: model}

This section contains a description of the proposed model.  The following sections, Sections~\ref{sec: sequences}-\ref{sec: parsing}, contain examples.

The representation I envision for a set of related concepts consists of three main parts:  an \emph{Intuitive Knowledge Structure (IKS)}, which stores intuitive concepts and their informal relationships, a \emph{Lexicon} that stores symbols and some of their basic attributes, and a \emph{Symbolic Knowledge Structure (SKS)}, which stores symbols and their formal relationships. 
The SKS connects to the Lexicon and the IKS.
Also, both the SKS and the IKS connect to input and output facilities, via special input and output neurons.

In addition, I assume that we have a limited amount of \emph{Working Memory}, consisting of neurons that can ``point to'' certain particular symbols or intuitive concepts that the brain is currently focusing on.
These Working Memory neurons may represent special ``roles'' being played by the symbols or concepts, such as the ``current number'' in a series of numbers, a partial result in a calculation, or the subject in a sentence.

The IKS contains entities (neurons, or collections of neurons) that represent intuitive concepts.  Reasoning using the IKS is intuitive reasoning, which I might not even call ``reasoning''---it is more like just following associations. The SKS contains entities that represent symbols.  Reasoning using the SKS is logical, rather precise reasoning.

I believe that the IKS is present in nearly all animals, not just humans.  The Lexicon and the SKS, however, seem to be mostly characteristic of humans.  Some research indicates that some apes may have rudimentary versions of these two structures, allowing them to process some symbols and perhaps perform some elementary symbolic manipulations (for example, see~\cite{Savage-RumbaughMSBWRB}).  
But in general, these structures are reserved for humans.

In the rest of this section, I sketch how we might model the various components formally.
My starting point is the synchronous, stochastic Spiking Neural Network (SNN) model of Lynch and Musco~\cite{LynchMusco-arxiv21}.
I expect that we will need to augment this model to include recent history, learning readiness, and other features.

\subsection{The Intuitive Knowledge Structure}

I begin by describing the Intuitive Knowledge Structure.  
In overview, the IKS consists of a collection of ``concept neurons'' and some directed connections between them.  The neurons and the connections may be unreliable. 

It seems natural to model the IKS at two different levels of abstraction, where we think of the lower level as ``implementing'' the higher level. 
The higher level consists of individual neurons, with a single neuron representing each concept. Some pairs of neurons are connected by directed edges.  
The lower level may allow multiple neurons for the same concept.  
The use of multiple neurons may provide redundancy and increase reliability---so a lower-level model based on less reliable neurons can ``implement'' a higher-level model with more reliable neurons.
I consider the question of whether the sets of neurons for different concepts can overlap, below.

\paragraph{Higher-level model:}
The higher-level model for the IKS is based on a weighted directed graph whose nodes correspond to concept neurons. 
This graph serves as the basis for a neural network model, similar to the one in~\cite{LynchMusco-arxiv21}:  at each round, each neuron exhibits firing behavior based on its total incoming potential, as described in~\cite{LynchMusco-arxiv21}.  This incoming potential is the sum of the weights of the incoming edges from currently-firing incoming neighbors.
%
The graph is dynamic: it may change over time as concepts get learned, get seen together, get seen in apparent causal relationships, and so on. 

This IKS model may include some unreliability, both in neurons and in connections.
To model unreliability in the neurons, note that the sigmoid function that is used in translating incoming potential to firing probability (see~\cite{LynchMusco-arxiv21}) yields stochastic behavior; shallower sigmoid functions yield more uncertainty.
We might also associate a failure probability with each neuron; failure of a neuron might mean that it doesn't fire even if its incoming potential indicates that it should, or fires even if it is not supposed to.
To model unreliability in connections, we might introduce a probability distribution for the potential conveyed on each edge, rather than using a fixed value.  We might also associate a failure probability with each edge; failure of an edge might mean that it doesn't convey potential, or conveys a different potential from what it should.
We might consider other failure modes, temporary or permanent.

I imagine that the IKS is constructed using operations like those presented by Valiant~\cite{Valiant}, and Papadimitriou and Vempals~\cite{PapadimitriouVempala}, notably the Join operation, also known as Merge.
In this work, performing a Join/Merge involves creating an entirely new concept that is built from other concepts, that is, a hierarchical concept.
Another operation is Associate, also known as Link.
These operations can be used to construct the IKS, and help to determine its structure.
Other structural relationships that might appear in the IKS include causality and temporal predictions.

For some purposes, we will probably want additional structure in the IKS, besides just neurons and directed edges.  For example, we might want to include the information that one concept neuron is a ``child'' of another concept neuron.  That is, we might classify some of the directed edges as \emph{parent} edges or \emph{child} edges, as in~\cite{LynchMallmannTrenn21}.
For a sequence, we might have \emph{successor} edges and \emph{predecessor} edges.
For a sentence, we might have \emph{subject} edges and \emph{predicate} edges.
Other labels might be appropriate for other kinds of relationships.
We also might want multiple edges, with different labels, between the same pairs of neurons.
Thus, in general, our model will be not just a weighted directed graph, but a labeled, weighted, directed multigraph.

Whatever type of graph we use, we will need to develop a Spiking Neural Network model based on it.  This will require extending the model of~\cite{LynchMusco-arxiv21}.\footnote{It is straightforward to extend that model to include multiple edges.  For edge labels, we may need other machinery.  For example, we might want to have external signals that turn all edges with certain labels on, or turn them all off.}  

We can consider several general kinds of operations to be performed on the IKS, with varying latencies.
\begin{enumerate}
    \item
    \emph{Direct recognition of concepts:}
    When an input is presented for one of the concepts in the IKS (in whatever form we allow for inputs, see Section~\ref{sec: model: io}), it can cause a particular neuron, or neurons, to fire.  This should happen very fast, in milliseconds.  This might also trigger some outputs (in whatever form we allow for outputs, see Section~\ref{sec: model: io}).
    
    This kind of operation corresponds to recognizing something directly, and responding in some automatic way.  This is ``fast thinking'', in Kahneman's terminology~\cite{Kahneman}.
    
    \item
    \emph{Indirect recognition of concepts:}
    An input is presented, but instead of causing immediate recognition, it triggers a cascade of neurons firing.  This might result in an output.  This overall process should still be fast, because each link is fast, as long as
    we are not talking about too long a chain of links.
    
    This corresponds to producing a collection of associations for the input concept.  Probably this is still ``fast thinking'', in Kahneman's sense, since it is automatic, and each step of the cascade is very fast.
    
    The behavior the network exhibits in such a cascade is as follows:  Input of some sort arrives, triggering some particular concept neurons to fire.  These cause potential to flow to their outgoing neighbor concept neurons.  If these receive enough potential, they fire with some probability, based on their sigmoid firing function (and possibly other uncertainty).  Thus, the effect of inputs is somewhat unpredictable, based on how this uncertainty is resolved.  The overall result from the input is a somewhat unpredictable cascade of firing of related concepts.

    \item
    \emph{Learning of concepts or relationships:}
    Some concept or relationship inputs are presented, causing new concepts or new relationships to be learned.  Also, already-learned concepts and relationships could have their representations strengthened when they are presented again.  
    
    Learning a new concept should involve associating the new concept with an unused internal neuron, plus strengthening the weights of edges that connect the input neurons encoding the concept to the internal representing neuron.  
    This process might use a Winner-Take-All mechanism to select a suitable unused neuron, as in~\cite{LynchMallmannTrenn21}.
    Reinforcing a concept might involve strengthening these edge weights.
    
    Learning a new relationship between two concepts, or strengthening an existing relationship, should involve increasing the weights of edges that connect the concepts to each other.
    
    These learning operations should be fairly slow---seconds, or minutes.
    They use some form of Hebbian learning, such as Oja's rule~\cite{Oja}.
\end{enumerate}

So far, the structure and behavior of this model appear to correspond well to Valiant's work~\cite{Valiant}.

What about outputs?
Among the concept neurons in the IKS, we can identify some as output neurons, which represent certain decisions.  For example, the decision may be an evaluation of a situation as ``good'', ``neutral'', ``bad'', or ``terrible''.
These neurons can also trigger emotional responses, like ``happy'', ``sad'', ``scared'', or ``angry'', probably by activating other neurons outside the IKS.

It should be possible to prove some simple theoretical latency bounds for the various kinds of operations on the IKS.  The recognition operations should be analyzed in terms of parameters representing time for a single round of execution, that is, for one step in the synchronous Lynch-Musco model, and the depth of a cascade of firing.
The learning time would depend on the time for a single round of execution, and on the particular learning rule that is used.
For example, Lynch and Mallmann-Trenn~\cite{LynchMallmannTrenn21}
analyzed the time required to learn a hierarchical concept, using a particular Oja-style learning rule.

\paragraph{Lower-level model:}
The lower-level graph is defined as a redundant version of a given higher-level graph.
Each concept $c$ represented in the higher-level graph has some number $m$ of neurons in the lower-level graph; to make things simple here, we assume the same number $m$ of neurons for every concept.
The edges are replicated, connecting every neuron representing one concept $c_1$ to every neuron representing another concept $c_2$ if and only if the unique representing neurons for $c_1$ and $c_2$ are connected in the higher-level graph.\footnote{This could be weakened to allow somewhat less connectivity.}
As for the higher-level model, this model may be dynamic, and may include some unreliability.
As before, we will want to allow multi-edges and labels, and we will need a neural network model based on this graph model.

The lower-level model supports the same kinds of operations as the higher-level model.
The direct and indirect recognition operations should work in essentially the same way as before, except that now the inputs will trigger multiple representing neurons to fire, propagation of firing will involve all the replicas, and the final output neurons will receive their potential from multiple internal neurons.
The learning operations will also involve replication, in that the same learning rule will be applied on all the edges that correspond to a single higher-level edge.  However, since the behavior of each individual neuron is stochastic, the behavior of the different replicas for the same concept will tend to diverge as the cascade proceeds.

An intriguing question is whether we should allow the sets of neurons representing different concepts to overlap.
Valiant, and Papadimitriou and Vempala, allow such overlap; in particular, Papadimitriou and Vempala analyze overlap in detail, in their work on computing with assemblies of neurons~\cite{PapadimitriouVempala}.
Based on their simulations, Papadimitriou and Vempala observe that the set of neurons that represent a particular concept may shift over time, in response to presentations of the concept in relationship with other concepts.

An alternative hypothesis that seems consistent with the results in~\cite{PapadimitriouVempala} is that the set of representing neurons for each concept $c$ is fixed, and does not change over time.
However, instead of observing the firing of a large set of representing neurons for a concept $c$ that shifts over time, we might observe the firing of a large set of neurons that participate in a multi-step firing cascade, initiated by the firing of a smaller, fixed set of actual representing neurons for $c$.
When two concepts are presented together, the connections between them strengthen, because of Hebbian learning.  Then afterwards, presentation of one of the concepts is likely to trigger firing of the representing neurons of the other, as part of a firing cascade.  This might make it look like $c$ has a larger, changing set of representing neurons, whereas actually, it has a smaller, fixed set of neurons, with changing edge weights. 

\paragraph{Relationship between the two levels:}
The low-level model adds redundancy to the high-level model, but otherwise operates similarly.  The key difference should be an increase in reliability, where many less-reliable neurons and edges at the lower level can be used to simulate more-reliable neurons and edges at the higher level.  This should lead to increased overall reliability. 

Once we have formal definitions of both the higher-level and lower-level models, we should describe a formal correspondence between the two system models and use it to prove that the less-reliable lower-level IKS model correctly simulates the more-reliable higher-level model.
The correspondence will define formal relationships between the states of the two system models, and between their transitions.
Since the systems are probabilistic, we would need to consider simulations specifically designed for probabilistic systems, such as the ones studied by Canetti, Cheung, et al.~\cite{CCKLLPS-jcss}, 
and Lynch and Segala~\cite{SL95}.

\subsection{The Lexicon}

The Lexicon stores symbols, along with some associated attributes, in a way that supports fast random access.  ``Random access'' here simply means that the firing of certain external neurons directly triggers the firing of neurons in the Lexicon, with no complex access mechanism involved.
A substantial Lexicon seems to be characteristic of humans, and does not seem to be present in other animals.

Formally, the Lexicon contains one (or perhaps a small number) of neurons for each symbol that it stores.  Input encoding a symbol is presented somehow, from outside the Lexicon, and triggers firing of the corresponding neurons in the Lexicon.  This input may come from several possible sources:  the outside world through vision, listening, or reading; corresponding symbol neurons in the SKS; or corresponding informal concept neurons in the IKS.
It is also plausible that the input from the outside world might not trigger the neurons in the Lexicon directly, but just indirectly, though symbol neurons in the SKS or IKS; these interactions remain to be defined.

In the human brain, I assume the Lexicon could be implemented in the neocortex.
This is consistent with the fact that humans have large and highly developed neocortexes, and also can store a very large number of symbols.
Also, the neocortex has a fairly uniform structure, which is consistent with the needs of a large random-access store.

\subsection{The Symbolic Knowledge Structure}

The symbolic knowledge structure (SKS) consists of a collection of ``symbol neurons'' and some directed connections between them.  The neurons and connections should be more reliable than those in the IKS.  Reliability for neurons may mean sharp thresholds (or at least, extremely steep sigmoid functions), plus no other types of failures.
Reliability for connections means that potential is conveyed accurately.
As for the IKS, we model the SKS at two levels of abstraction.

\paragraph{Higher-level model:}
Abstractly, the SKS is another labeled, weighted, directed multigraph.
The nodes of this digraph are ``symbol neurons'', which are separate from the corresponding symbol neurons in the Lexicon.
There are reliable edges between the corresponding symbol neurons in the SKS and Lexicon, in both directions.
The graph is dynamic, changing in response to presentation of new concepts and combinations of concepts.

The SKS can be constructed using operations like those used for the IKS, such as  Join/Merge and Associate/Link.  
However, more interesting special-case structures (such as sequences, parse trees, mathematical proofs,...) can also be built in the SKS, under purposeful control from the outside.  This may use Working Memory.

The SKS is also connected to input mechanisms (visual, auditory) and output mechanisms.  Each symbol neuron can be triggered by symbolic inputs, like hearing a particular word spoken.  Likewise, the neuron can trigger symbolic outputs, like saying the word.
Another important input/output mechanism for symbols involves reading and writing.
Reading and writing differ from speech input/output in that they have the pleasant effect of expanding the SKS, increasing its size by utilizing paper or computer storage.
Thus, humans can write down a long list of items, or a complicated math proof, or a very large parse tree.  These symbolic structures can be much larger than what humans can remember, or manipulate, in their internal SKS structures.\footnote{
Another example of expanding the SKS is the process of ``counting on our fingers''.  Mostly everyone over the age of five knows a correspondence between our fingers and the numbers one through ten, and can use that for counting items in the real world. 
We can also use our fingers to help us count symbols in a sequence that is memorized in the SKS.}

Furthermore, and crucially, the SKS is directly connected to the IKS.
Specifically, each symbol neuron in the SKS connects to one or more related concept neurons in the IKS.
There should be a strong and reliable connection from each symbol neuron to each of its corresponding intuitive concept neurons, and vice versa.\footnote{
It seems clear that humans can generally move quickly from a symbol neuron to any of its corresponding intuitive concept neurons.   The reverse is also usually true, but there are exceptions---when you just can't think of the name for some concept that you ``know''!}

Again, we can consider several kinds of operations to be performed on the SKS, with different latencies:
\begin{enumerate}
    \item 
    \emph{Direct recognition of symbols:}
    As for the IKS, when an input is presented for one of the symbols in the SKS, it causes a particular neuron, or neurons, to fire.  As for the IKS, this should happen very fast, in milliseconds.  This might also trigger symbolic outputs.  In addition, the firing of a symbol neuron in the SKS will trigger firing of the corresponding neuron in the Lexicon.
    
    As before, this corresponds to recognizing a symbol quickly and directly, and perhaps responding in some automatic way.  As for the IKS, this part of the usage of the SKS corresponds to Kahneman's fast thinking. 
    An example here is direct translation, such as hearing a spoken word and outputting a written version.
    
    \item
    \emph{Multistep symbolic computation:}
    An input is presented, and it triggers some type of symbolic computation in the SKS.  For example, a number $k$ may be input, and the number $k+2$ should be output.  Or two numbers may be input and their sum should be output.  
    Another example, which I will explore in Sections~\ref{sec: sequences} and \ref{sec: diseases}, involves inputting a number $k$ and outputting the $k$th element in a sequence that is memorized within the SKS.
    
    Another example, which I will discuss in Section~\ref{sec: parsing}, involves inputting a stream of words, and having the SKS determine whether or not the words form a syntactically correct sentence having a predetermined simple structure, say, a noun (the subject), followed by a transitive verb, followed by another noun (the object). The SKS might also output a representation of the sentence structure, such as a parse tree.  
    More generally, the SKS might determine whether the sentence fits any of a predetermined finite set of sentence structures, or fits any correct sentence structure.
    Other examples of multistep symbolic computation might include solving algebraic equations.
    
    The latency required for such operations varies, of course.  
    For recognizing simple sentences, the response should be very fast:  people parse simple sentences almost immediately after the input has been fully presented.  
    But processing even a simple sentence isn't trivial:  it requires finding the words in the SKS and the Lexicon, and fitting them into some representation of the sentence structure, such as a parse tree.
    Doing all of this fast might require that the sentence structure be built into the SKS ahead of time, so that what remains to be done is just fitting the arriving words into their proper places.
%
    For more complicated sentences, time is also required for determining the sentence structure.
    
    Counting, or more elaborate arithmetic calculations, will give a slower response, on the order of seconds.  These require multi-step calculations, and following paths in the SKS.  They also require attention, and purposeful focus on particular parts of the computation.  This focus will involve the use of Working Memory (see below), and perhaps some form of Winner-Take-All mechanism~\cite{LynchMP-arxiv19}.  
    This type of computation should correspond to Kahneman's slow thinking.
    
    All of these multi-step computations in the IKS should be precise and reliable, in contrast to multi-step computations in the IKS.
    
    \item
    \emph{Computation involving both the SKS and the IKS:}
    The simplest case involves computation in the SKS, followed by triggering of neurons in the IKS and cascades of firing in the IKS.  Our examples in Sections~\ref{sec: sequences}-\ref{sec: parsing} are all of this type.  More complicated cases would involve more elaborate interactions between the SKS and IKS.  
    
    In the simplest case, input is presented, and it triggers some symbolic calculation in the SKS.  The firing of some of the symbol neurons in the SKS triggers the firing of their corresponding concept neurons in the IKS.  These, in turn, trigger a cascade of firing in the IKS, eventually resulting in some output. 
    For example, inputting the number $k$ triggers firing of a symbol neuron for the symbol $7$ in the SKS, which in turn can trigger firing of some concept neuron for that number.  The IKS then produces a cascade of associations for that concept; for instance, for the number $7$, we can get the concept of seven deadly sins, seven samurai, or seven dwarfs.  An output might be a rudimentary mental picture of one of these.
    The latency for this sort of operation is only a bit more than that for the multistep calculations within the SKS, because the computation in the IKS is fast.
    
    In a more complicated case, the semantics and intuitive associations for words, which arise in the IKS, can help the SKS in determining a sensible parse tree for a given input sentence.
    Another interesting interaction arises in solving a math problem:  Here, the problem is input to the IKS, which uses its intuitive reasoning to produce ideas for solving the problem.  These trigger the SKS to perform some formal calculations, producing partial results. These partial results then serve as new inputs to the IKS, which uses them to produce more ideas, and so on.
    The latency for such complex operations is essentially the sum of all the computation times in the IKS and SKS during all the phases of the interaction.

    \item
    \emph{Learning of concepts or relationships:}  
    Analogously to the IKS, the SKS can be modified by adding new symbol nodes and new connections, and by strengthening existing connections.
    Learning a new symbol should be as in the IKS, using some generic process for selecting an unused neuron.  Again, this may use a Winner-Take-All mechanism.
    
    However, adding or strengthening connections between neurons should be different in the SKS from in the IKS: instead of using incremental Hebbian rules to strengthen edges that are frequently used, the SKS may make bigger, ``all-at-once'' increases in weights.  This makes sense in view of the assumption that the changes in the SKS are produced purposefully, triggered by some external stimulus, rather than emerging accidentally as a result of random experiences.
%
    Making larger changes should lead to faster learning than an incremental Hebbian learning process.  
    
\end{enumerate}

\paragraph{Lower level model:}
At a lower level of abstraction, the SKS can be represented by another labeled, weighted, directed multi-graph, in which each concept has some number $m$ of representing neurons, with all such sets disjoint.  Edges can be similarly replicated.  All this is as for the IKS, but we do not allow overlapping sets of neurons here. This ensures that the lower-level model for the SKS maintains clear separation between the representations of different symbols.  This seems to be necessary for reliable computation.

As for the IKS, the lower-level SKS model supports the same operations as the higher-level model. 
%
%
As for the IKS, the redundancy in the lower-level SKS model should yield greater reliability in the higher-level SKS model.   

\paragraph{Comparison between processing in the SKS and IKS:}
Processing in the SKS models is somewhat different from processing in their IKS counterparts.
For one thing, I assume that the individual elements of the SKS models are more reliable than those in the IKS models, which implies that processing should be generally more reliable. 
But also, there is a difference between the kinds of computations that are carried out in these two structures.  
In the IKS, when neurons fire, they trigger neighboring neurons to fire, according to general stochastic firing rules. On the other hand, the SKS is subject to more external control, which chooses which symbols to focus on and guides the computation step-by-step.  An external mechanism probably helps in keeping track of intermediate results of the computation---see Section~\ref{sec: model: wm} for a discussion of Working Memory.
Thus, the SKS can carry out precise, systematic computations, such as arithmetic calculations, whereas the IKS just produces general associations.

\subsection{Input and output}
\label{sec: model: io}

Both the Symbolic and Intuitive Knowledge Structures accept inputs and produce outputs.

\paragraph{Inputs:}
In both the SKS and IKS, inputs arrive via special input neurons.
For the SKS, these inputs encode symbols, and for the IKS, they encode intuitive concepts.
The encoding scheme may depend on the type of input.  
For example, number inputs to the SKS could be input in unary or binary.  
For the IKS, we could have a unique input neuron for each intuitive concept, or a set of neurons (possibly with overlap between the sets for different concepts),
or even a binary code.
The inputs can be visual, auditory, written, olfactory, or tactile. 
Visual input might be provided using a feature vector, as in machine learning.

\paragraph{Outputs:}
For the SKS, the outputs take the form of symbolic output, which might be drawn, written, spoken, or gestured.
For the IKS, the outputs may be general impressions, such as an assessment of whether something is good, bad, or terrible, or it could be an emotional response.  The SKS outputs could also be triggers for motor actions.

\subsection{Working memory}
\label{sec: model: wm}

The final component of the knowledge system is something I call the Working Memory.  This is essential for computation in the SKS.\footnote{ I think it is very likely that a rudimentary type of Working Memory could also be used with the IKS, allowing the brain to focus attention on some intuitive concept without naming it with a symbol.  But I will not discuss that here.}
The Working Memory contains a limited number of neurons that are used to keep track of intermediate results in an SKS computation. 
I think of the Working Memory neurons as representing ``roles'' in a computation, to be filled by specific numbers, words, or other symbols during computation.  
It might be useful to  think of them as pointers into the SKS structure.

For example, in a counting process, one Working Memory neuron might keep track of the number $k$ that the system is up to in the count.
If the system is counting symbols in a memorized sequence, another Working Memory neuron might keep track of the current symbol in the sequence.
If the system is counting up to some goal number $g$ (for example, because it is counting up to the $g$th  element in a memorized sequence), then a third Working Memory neuron might indicate the goal. 
For parsing a simple sentence, the Working Memory might keep track of the subject, predicate, and object in the sentence.  

\paragraph{Implementing Working Memory:}
An important question is exactly how we should implement the Working Memory in a Spiking Neural Network model, specifically, how to represent the idea that a ``role neuron'' in Working Memory ``points to'' a particular symbol neuron in the SKS.
We can't actually store a pointer in a neuron, as we do in a programming system, so we need to emulate this connection somehow.

One idea is to say that a role neuron points to a symbol neuron at a certain time $t$ if and only if the two neurons both fire at that same time $t$.
We might interpret this as meaning that the Working Memory is paying attention to, or focusing on, a particular symbol, at time $t$.
In this case, we might say that the role neuron is ``bound to'' its current corresponding symbol neuron.
We might strengthen this condition by requiring that the role neuron and symbol neuron fire together for some number of consecutive times.
Perhaps other conditions will need to be added.

But this has an obvious problem:  the same role neuron can fire together with two (or more) symbol neurons.  So according to this definition, that would mean that one role neuron is pointing to more than one neuron at the same time.  This does not seem too sensible if the binding between role neurons and symbol neurons is supposed to represent focus.
One way out might be to have the system maintain an invariant that prohibits such ``split attention''.  Implementing such an invariant might require a Winner-Take-All mechanism.
Other technical issues arise, for example:
\begin{itemize}
    \item 
    Suppose that we have two role neurons, each pointing to its own symbol neuron.  In that case, we could allow all four neurons to fire together, but this seems confusing.  It might be cleaner to separate the firing for the different bindings, for example, alternating the firing of the two pairs of neurons according to some simple pattern.\footnote{
    Implementing such alternation might involve using some neurons as memories, along the lines of work by Hitron et al.~\cite{HitronLMP20}.}
    \item 
    What about two role neurons that point to the same symbol neuron?  In that case, we could allow all three neurons to fire together.  But again, an alternation discipline would probably be cleaner here:  at some times, one role neuron fires along with the symbol neuron, and at other times, the other role neuron fires along with the same symbol neuron.
\end{itemize}

Just having a role neuron and a symbol neuron firing together, in some pattern, doesn't give us very much.  We have to understand how this synchronized firing helps in performing computation in the SKS, for example, in counting.  Some preliminary thoughts about this appear in Section~\ref{sec: diseases}.  For now, we just leave this as:

\paragraph{Open question 1:}  Devise a concrete SNN mechanism for implementing a Working Memory, based on synchronizing role neurons and symbol neurons.  Give specific examples of how this mechanism can be used in symbolic reasoning.

\subsection{Humans vs. animals}

Considering two knowledge structures, intuitive and symbolic, may help us to understand differences in thinking between humans and other animals.
Non-human primates and other mammals have very well developed intuitive knowledge structures, in some ways better than humans' structures, at least for the types of concepts that are important for the animals' survival and comfort.
I imagine that this structure is very similar to intuitive knowledge structure in humans.

Thus, like a human's IKS, an animal's IKS is a labeled, weighted, directed multigraph, with unreliable nodes and edges. 
It connects to visual, auditory, and olfactory input neurons, and decision, emotion, an motor action output neurons.
It receives input signals, which trigger a cascade of firings, which eventually trigger output behavior.  It also supports learning, in that inputs can cause modifications to this structure, using Hebbian-style rules.

However, I believe that animals have only a small, rudimentary symbolic knowledge structure (SKS), if any.
Some animals (the most advanced apes) seem to be able to acquire a small vocabulary of symbols and use them in simple ways~\cite{Savage-RumbaughMSBWRB,Savage-RumbaughLewin}.
These should be able to fit into a small Lexicon, and might be processed in simple ways using a rudimentary SKS.  
We might say that animals that exhibit some basic ability to use symbols have a limited form of this fundamentally human characteristic.
I don't consider that to be a contradiction, only an indication of the beginnings of a human capability in some higher primates.

Can we match the theoretical IKS and SKS structures with actual areas of the human brain?
For example, can we map the SKS to some part of the left brain and the IKS to some part of the right brain?  
Also, Friederici~\cite{Friederici} provides evidence that certain connections between brain areas involved in language processing are much stronger in adult humans than in animals (or infant humans). Can we match these real connections with theoretical connections between the SKS and IKS?

\paragraph{Language processing:}
What is the relationship between the SKS/IKS dichotomy and language processing?
I think it is natural to regard a parse tree, and more generally, the collection of parse trees representing the possible structures of sentences in a language, as a symbolic knowledge structure.
Let's assume for now that this collection of possible parse trees is already memorized in the SKS, and not worry now about how it got there, whether by learning or evolution.

Then what happens when a person hears (or reads) a sentence?
They match up the words of the sentence with leaves in the parse trees in the SKS, and larger parts of the sentence with internal nodes in the parse trees.
Somehow, this happens very quickly, at least for sentences with simple structure (such as 3-word (noun, transitive verb, noun) sentences).
The process of parsing a simple sentence may use Working Memory to describe the ``roles'' of the major parts of the sentence, such as subject, predicate, and object.
It may also use semantic knowledge from the IKS, to help in disambiguating the sentence.
Viewed in this way, the problem of parsing a sentence is a special case of symbolic computation in the SKS, possibly with some contribution from intuitive processing in the IKS.

So then, what is the most important difference between human and animal capabilities with respect to language processing?
I believe it is just that humans have a well-developed SKS that is capable of representing and manipulating formal relationships among linguistic symbols. 

Berwick and Chomsky's theory of linguistic processing~\cite{BerwickChomsky} reaches a rather different conclusion.
They propose that the key difference is the ability of humans to build arbitrarily complex hierarchical mental structures, such as complex sentences, using Merge operations and recursive grammar rules. 

I don't think that can be exactly right.  I have three comments.
First, the Join/Merge operation, as studied, for example, by Valiant, seems to be applicable for intuitive concepts in the IKS, at least as much as for symbols in the SKS.
Since the IKS is present in both humans and animals, it is likely that animals can perform simple intuitive Merges.  
In fact, I suspect that there is little difference between the abilities of humans and higher animals to construct intuitive hierarchies in the IKS.

Second, when we talk about recursive grammar rules, we are talking about sophisticated symbol manipulations, which humans can do and animals cannot.
Then we are in the realm of symbolic reasoning, which involves the SKS.

And third, I don't think that humans can actually construct arbitrarily complex hierarchies in their SKSs.
%
%
%
For instance, it seems to me that a human can keep track of only three or four levels of structural depth in a sentence's parse tree.
The SKS in the human brain probably has some graph-theoretic limitations, which impose limits on the size and depth of the linguistic structures that can be represented.
Also, perhaps understanding a sentence with very complex structure requires focusing on a large number of symbols at once---more than can be handled in Working Memory.\footnote{Of course, if we expand the SKS and Working Memory by writing things down on paper, as humans can do, then much more complex hierarchies could be represented. But this is not a fair comparison---let's stick to the internal capabilities of the brain.}

At any rate, it seems to me that the important distinction in linguistic abilities is simply that humans can do sophisticated symbolic processing, whereas animals cannot.

\vspace{.5cm}
In Sections~\ref{sec: sequences}-\ref{sec: parsing}, I give two examples to illustrate the ideas of Section~\ref{sec: model}.
We focus on the case where the structures of interest are already learned, and mention learning only later, in Section~\ref{sec: learning}.
Our examples involve traversing sequences of symbols and parsing simple sentences.

\section{Sequences}
\label{sec: sequences}

To make the ideas of Section~\ref{sec: model} more concrete, I define here here a simplified, special case of a symbolic knowledge structure, namely, a memorized finite sequence of distinct symbols. 
Using this SKS, one might compute, for example, the symbol that appears in a particular position in the sequence.
Following this computation with further computation using an intuitive knowledge structure might produce some reactions related to a concept denoted by this symbol.

\subsection{Representation}

Let's consider a finite sequence of $k$ distinct symbols, $s_1, s_2,...,s_k$.  Let $S$ denote the set of symbols in the sequence.

We assume that the SKS is a linear directed graph, with nodes (= neurons) corresponding to the $k$ symbols.
Denote the node for symbol $s$ by $rep(s)$.
These nodes are connected in increasing sequential order, where each $rep(s_i)$ has a directed edge to $rep(s_{i+1})$.
In addition, each $rep$ node has a self-loop.
Weights on all these edges are positive, so that firing of one $rep$ node at one time $t$ encourages the firing of the next $rep$ node in the sequence, and also encourages its own firing, at time $t+1$.
We assume that the nodes are threshold elements or have their firing determined by very steep sigmoids, and all components behave reliably.

The IKS contains one or more intuitive concept neurons for each symbol in the SKS.
These connect to other concepts using positive weight edges, in arbitrary ways.  Concept neurons for the successive elements of $S$ are not connected to each other in a systematic pattern.
%
The neurons may use shallow sigmoids, and the components need not behave completely reliably.

For each symbol $s \in S$, we also have directed edges between $rep(s)$ and each intuitive concept neuron for $s$, in both directions.

\paragraph{Example:  The Greek alphabet and virus variants:}
We consider the sequence of 24 Greek letters, alpha, beta, ..., omega.
This can be memorized in the SKS, in the manner just described.

Each Greek letter is associated with numerous intuitive concepts, including mathematical notions, hurricanes, and most importantly for this paper, a variant of a certain aggressive virus.  Each of these intuitive concepts has a corresponding intuitive concept node in the IKS.

For example, consider virus variants.  
For this paper, we pretend that every Greek letter has exactly one associated virus variant.\footnote{For the COVID virus, that is not quite true:  we are currently not done with the list, but are stalled at omicron.
Also, someone decided to skip the letters nu and xi, for reasons that should be easy to guess.}
Loosely associated with a virus variant are many of its properties, including where it originated, how contagious it was, how serious an impact it had, approximately how many people caught it, whether it caused widespread business and and school shut-downs,  whether a lot of people died, etc. This is all informal understanding, and is represented by nodes and edges in the IKS.
These thoughts are all produced quickly, along the lines of Kahneman's fast thinking. 

In order to define abstract problems involving Greek letters and virus variants, we also include special output neurons in the IKS.
These can represent informal decisions about the virus variant, such as whether the variant was OK, bad, or terrible.
They can also represent emotional reactions, such as neutral, scary, or terrifying.

\subsection{Queries}

We define two natural queries for sequences, both of which involve counting symbols and evaluating associated intuitive concepts.
Answering them will involve a combination of symbolic and intuitive reasoning.
For these queries, we assume that each symbol neuron has exactly one corresponding intuitive concept neuron.  

\begin{enumerate}
    \item  
    The input is a ``goal number'' $g$ in the allowed range $1,...,k$, where $k$ is the length of the  sequence. The output should be a decision, or emotional reaction, derived from the firings in the IKS starting from $rep(s_{g})$.

    What is the ``correct'' response for this query?  This is defined by the structure---edges and weights---of the relevant part of the IKS, which can be arbitrary.  
    %
    According to the IKS execution rules, when the concept neuron $rep(s_{g})$ fires, it triggers a cascade of firing of other IKS neurons.  This yields a probability distribution on decision neurons, or emotional reaction neurons, that are reached by the cascade, by any particular time.
    We require that the output for the query follow this distribution.
    
    In the most interesting case, the IKS operation guarantees that, for some time $t$, with very high probability, the firing decision or emotional reaction neuron is unique, and moreover, that the firing pattern for decision/emotion neurons persists from time $t$ onward.
    In this case, the query response should also stabilize to a unique value by some time $t$, with high probability.

    \item  There are two inputs, a symbol $s$ in the set $S$ and a goal number $g$ in the range $\{1,...,k-i\}$, where $s = s_i$.  The output should be the correct decision or emotional reaction, derived from the firings in the IKS starting from $rep(s_{i+g})$.
    
    The correct response for this query is defined as in the first query, based on IKS behavior starting from $rep(s_{i+g})$.
\end{enumerate}

\paragraph{Answering the queries:}
The first issue that must be addressed, for processing queries like these, is the representation of their  inputs.  In this case, how does one input a number in $\{1,...,k\}$?  How does one input a symbol $s$?
I propose that, for now, we consider the simplest possibilities:
\begin{itemize}
\item
For numbers in $\{1,...,k\}$, use simple unary representation.  Assume we have a dedicated input neuron for each number in this range, which is triggered to fire by some external stimulus, in order to input the number.
\item
For symbols in $S$, we could allocate one input neuron for each symbol.  
\end{itemize}

As it happens, the processing of these two queries can be broken down into sequential phases:  provide symbolic inputs, use symbolic computation to identify a particular symbol neuron in the SKS, move to the corresponding intuitive concept neuron in the IKS, and then let the cascade in the IKS produce the final result.  One could also consider queries whose processing involves more elaborate interactions between the two structures; examples of this sort might arise, for example, in parsing and understanding sentences.
But for now, let's focus on these simple Queries 1 and 2, with their stylized one-way communication pattern between the SKS and IKS.

During the symbolic processing phase of these queries, we will also use the Working Memory, to keep track of particular symbols that the computation is currently focusing on.

So how might the answers to Queries 1 and 2 be computed?  For Query 1, we will count up to the input goal number $g$, moving step-by-step through a sequence of neurons representing consecutive positive integers.
As we do this, for each step, we also move one step in the sequential SKS.  
When we reach the number $g$, we must recognize that we have reached it, and identify the associated intuitive concept neuron.  The firing of this neuron starts the cascade that eventually produces the ouput.  It remains to express this strategy as a formal SNN; a start at this appears in Section~\ref{sec: diseases}.

Query 2 will be answered similarly.  This time, we start at the symbol neuron in the SKS that is indicated by the symbol input.  Again, we traverse the SKS while counting up to the input number $g$, then continue processing in the IKS.
Thus, both of these queries involve simple linear traversals of the SKS, while counting the
number of steps that have been traversed.

\paragraph{Example:  The Greek alphabet and virus variants:}
For Query 1, we have $k=24$, so the goal number input is any number $g \in \{ 1,\ldots,24 \}$.  The output is an assessment of how bad the virus variant was that was associated with the Greek letter at position $g$ in the Greek alphabet.  For example, input $4$ gives the delta variant, which might produce output ``terrible'' based on the number of cases and deaths.
Similarly, for Query 2, we ask how bad the virus variant was that is associated with the Greek letter in position $g$ after the variant symbolized by $s$.


\paragraph{Complexity analysis:}
When we define formal SNN algorithms for answering these queries, we would like some way of measuring their computational complexity.
I focus here on latency, measured in terms of the number of basic steps in the SNN model.

The strategies outlined above involve computation in the SKS followed by computation in the IKS.
In this special case, we can obtain a bound on latency simply by adding the time bounds for these two phases of computation (plus one extra step for the time from when the final symbol neuron fires to when its corresponding intuitive concept neuron fires).

The time for computation in the IKS is the time from when the initial intuitive concept neuron fires until a decision or emotion neuron fires persistently.
This corresponds to the time to follow some number of links in the IKS, plus perhaps some time to suppress all but one candidate decision.  We don't know the number of links, and it is not clear yet how many candidate decisions there will be, nor exactly how one decision will be selected.  So for now, we encapsulate all this in a parameter, $t_{iks}$.
In any case, this should be a small number, qualifying as fast thinking in Kahneman's sense.

The time for computation in the SKS will be the time to follow $g$ links in the SKS, focusing on one link at a time.  It is not clear how long it will take to follow a single link, so for now, we make this another parameter, $t_{sks}$.
Then the total time for each of the queries would be roughly $g t_{sks} + t_{iks} + 1$.  

In Section~\ref{sec: diseases}, we give some technical details for possible mechanisms for answering these queries.  The reader may want to skip that section for now, and move on to our other example, in Section~\ref{sec: parsing}.

\section{More Details for Reasoning about Sequences}
\label{sec: diseases}

Here I consider theoretical mechanisms that might be involved in answering queries like those in Section~\ref{sec: sequences}.  

We need a mechanism for counting up to the $g$th symbol in the memorized sequence.
I assume that Working Memory includes three neurons, a \emph{current-number} neuron that keeps track of a natural number $k$ representing where we are up to in the count, a \emph{current-symbol} neuron that keeps track of the corresponding symbol in the sequence, and a \emph{goal} neuron that holds the goal number $g$.  These neurons represent three distinct ``roles'', which will, during the computation, synchronize with particular symbol neurons that fill these roles.

We must pin down exactly how a role neuron synchronizes with a symbol neuron.  I described some ideas for this in Section~\ref{sec: model: wm}, but left the specifics as an open question.
Basically the role neuron and a corresponding symbol neuron fire together, though questions remain about exactly what synchronization conditions they satisfy. 
Another issue is how the system produces this synchronized firing.
This may require that the graph contain directed edges in both directions, between the role neuron and all the possible symbol neurons in the SKS that might potentially fill this role.
These edges can contribute potential between the two neurons so that once they are both firing, each can help the other to continue firing.

In the rest of this section, I elaborate on some issues involved in answering the queries.  Section~\ref{sec: increment-count} describes the core of the implementation: a mechanism for incrementing the count by $1$ while moving one step in the sequence of symbols.  Section~\ref{sec: start-end} describes how to start and end the counting process.  Section~\ref{sec: tie-together} discusses how to tie it all together to answer the queries.

\emph{Note:}  We have a small confusion in terminology here.
So far, I have been referring to the elements in the memorized sequence as ``symbols''.
But symbols are really a broader classification, including, for example, natural numbers.
In this section, I will be dealing with natural numbers as well as elements of the memorized sequence, and I will need to distinguish the two.
So for the rest of this section, I will use the word ``symbols'' generically, to indicate any kind of symbol, ``numbers'' for natural numbers, and ``letters'' for the symbols in the memorized sequence (as in the Greek alphabet example).  I will rename the \emph{current-symbol} neuron to be the \emph{current-letter} neuron. 

\subsection{Incrementing the count}
\label{sec: increment-count}

I begin with a sketch of a mechanism for incrementing the count by one.
Here I use the notation $rep(s_i)$ for the neuron for the $i$th letter (as I defined earlier), and also use $rep(i)$ for the neuron for number $i$.

Consider any fixed $i \in \{1,\ldots,k-1\}$.
The increment operation starts with the \emph{current-number} neuron in the Working Memory bound to neuron $rep(i)$, and the \emph{current-letter} neuron in the Working Memory bound to neuron $rep(s_i)$.
Specifically, we start with the \emph{current-number} and \emph{current-letter} neurons firing, along with the symbol neurons $rep(i)$ and $rep(s_i)$.
%
%
The result should be that the \emph{current-number} neuron gets bound to $rep(i+1)$, and the \emph{current-letter} neuron gets bound to $rep(s_{i+1})$.
That is, we should end with the \emph{current-number} and \emph{current-letter} neurons firing, along with the symbol neurons $rep(i+1)$ and $rep(s_{i+1})$.

We assume that the increment is triggered by some special \emph{next} input signal, produced by the firing of a special input neuron.

\paragraph{Incrementing numbers only:}
First, I ignore the letters and focus just on incrementing the numerical count.  I start with a static description of the neurons and connections.
We have a single \emph{current-number} neuron in the Working Memory and a sequence of number neurons in the SKS.  All are modeled as threshold gates.
The number neurons have threshold given by a parameter $h$; the threshold for the \emph{current-number} neuron is $0$.

The \emph{current-number} neuron has directed edges to all the number neurons, each having weight $c$.
Each number neuron has a directed edge to the next higher number neuron, with weight $s$.
Each number neuron also has a self-loop, with weight $l$.

In addition, we will have some incoming edges that provide external control of parts of the computation.  
Specifically, we have incoming edges to all the number neurons from a single external excitatory source, with weight $exc$, and likewise incoming edges to the number neurons from an external inhibitory source, with weight $inh$.
More formally, we define the following parameters:
\begin{itemize}
    \item $h$, the threshold for each number neuron.
    \item $cur$, the weight on the edge from the \emph{current-number} neuron to each number neuron.
    \item $l$, the weight on the self-loop for each number neuron.
    \item $s$, the weight on the edge from each number neuron to its successor number neuron.
    \item $s'$, a value somewhat less than $s$.
    \item $exc$, the incoming weight on the edges from an external excitatory source to all the number neurons.
    \item $inh$, the incoming (negative) weight on the edges from an external inhibitory neuron to all the number neurons.
\end{itemize}
\noindent
We require that the parameters satisfy the following inequalities:
\begin{enumerate}
    \item $h \leq cur + l$
    \item $h > s + cur$
    \item $h \leq exc + s + cur$
    \item $h > exc + cur$
    \item $h > cur + l + inh$
    \item $h \leq cur + l + inh + s'$
\end{enumerate}

Many values satisfy all of these inequalities, such as:
$cur = 0$, $l = 4$, $h = 3$, $s = 2$, $exc = 2$, $inh = -2$, $s' = 1$.
%

Now, how does this mechanism operate?
At time $0$, the \emph{current-number} neuron and number neuron $rep(i)$ are both firing.  
The subsequent behavior is completely determined by the structure of the network, the initial settings, and the arrival of external inputs.
(There is no stochasticity here, since the involved neurons are all deterministic threshold elements.)
The mechanism is triggered to start the increment process by the arrival of an excitatory input from an external source at all the number nodes.  This is supposed to start the firing of number neuron $rep(i+1)$.
After a small delay, an inhibitory input arrives at all the number nodes.  This is supposed to stop the firing of number neuron $rep(i)$.
Here, informally, are the main requirements on the mechanism's behavior:

\begin{itemize}
    \item 
    If the \emph{current-number} neuron and number neuron $rep(i)$, are firing, they continue firing without any external input.  If $rep(i)$ is firing, it doesn't trigger number neuron $rep(i+1)$ to fire in the absence of any other inputs to $rep(i+1)$.
    \item
    If an external excitatory input arrives at all the number neurons while number neuron $rep(i)$ is firing, the next number neuron, $rep(i+1)$, starts firing, while $rep(i)$ continues firing.  No other number neurons start firing.
    \item
    If both $rep(i)$ and $rep(i+1)$ are firing, and an inhibitory input arrives at all the number neurons, then $rep(i)$ stops firing, while $rep(i+1)$ keeps firing. 
\end{itemize}

These considerations motivate the inequalities on the parameters, as follows.  
Inequality 1 says that $h \leq cur + l$.
This says that the incoming potential from the \emph{current-number} neuron, plus the incoming potential from the self-loop on number neuron $rep(i)$, are enough to exceed $rep(i)$'s firing threshold.
That means that $rep(i)$ continues firing without any external input.

Inequality 2 says that $h > s + cur$.
This says that incoming potential from number neuron $rep(i)$, plus the incoming potential from the \emph{current-number} neuron, are not enough, on their own, to trigger firing of $rep(i+1)$.

Inequality 3 says that $h \leq exc + s + cur$.
This says that, when the incoming potential to number neuron $rep(i+1)$ from $rep(i)$, and the incoming potential from the \emph{current-number} neuron, are augmented by an external excitatory signal, the combination is enough to exceed $rep(i+1)$'s firing threshold, and thereby trigger it to fire.

Inequality 4 says that $h > ext + cur$.
This says that incoming potential from the external signal, when added to that from the \emph{current-number} neuron, is not enough to cause any other number neuron besides $rep(i+1)$ to start firing.

So far, we have argued the first two bullets above, which describe how number neuron $rep(i+1)$ starts firing.
The third bullet deals with stopping the firing of $rep(i)$.
First, Inequality 5 says that $h > cur + l + inh$.
This means that the incoming (negative) potential from an inhibitory source, when added to the potential from the \emph{current-number} neuron and the self-loop, is not enough to reach the threshold for $rep(i)$.  So this will have the effect of stopping the firing of $rep(i)$.

But now we must be careful, because we don't want the inhibition to also stop the firing of $rep(i+1)$.  When $rep(i)$ neuron stops firing, its contribution of potential to $rep(i+1)$ ceases. So that leaves only potential $cur$ from the \emph{current-number} neuron, plus $l$ from $rep(i+1)$'s self-loop, reduced by the inhibition amount $inh$.  Inequality 1 tells us that $h \leq cur + l$, but we don't have anything to tell us that $h \leq cur + l + inh$, which is what we need here.

One way out is to extend the model of~\cite{LynchMusco-arxiv21} 
by allowing the neurons to remember some history of incoming potential.  We have extended the model in this way before~\cite{SuCL-jour19}.
Essentially, we want to allow the potential contribution from $rep(i)$ to degrade over a short period of time (for the time during which the inhibitory input persists), rather than disappear immediately.
We could capture this by assuming that some residual potential contribution of magnitude $s' < s$ persists for at least as long as the interval when the inhibition is being applied.
Then Inequality 6, which says that $h \leq cur + l + inh + s'$, will tell us that $rep(i+1)$ has enough incoming potential to keep firing during the inhibition interval.
This seems rather delicate, but not necessarily unrealistic.

\paragraph{Incrementing numbers and letters:}
How can we augment the mechanism just defined to increment both the number and the corresponding letter?
Well, it seems that the same mechanism can work for both sequences, sequentially or in parallel.  For a parallel solution, we can send the same excitatory and inhibitory signals to both the number neurons and the letter neurons.
There should be no interaction between the two parts.
A sequential solution would involve separate excitatory and inhibitory signals for the two types of neurons, with alternating applications of the signals for the numbers and the letters.

\subsection{Starting and stopping the count}
\label{sec: start-end}

Next, I consider how to start and stop the count. 
Consider Query 1.

\paragraph{Starting:}
We need a way of triggering the firing of the first number neuron, $rep(1)$, and the first letter neuron, $rep(s_1)$.
Since these neurons have no incoming potential from predecessor neurons, we will use some special input to get them firing.
Namely, we assume an external excitatory neuron, with edges to both the \emph{current-number} and \emph{current-letter} neurons, with sufficiently high weight to meet their thresholds.
The external neuron also has edges to the number neuron $rep(1)$ and the letter neuron $rep(s_1)$.
These latter two edges have weight $s + exc$ instead of just $exc$ as in the incremental mechanism. The extra $s$ compensates for the fact that neurons $rep(1)$ and $rep(s_1)$ don't receive potential contributions of $s$ from their predecessors.

We also must set the goal to a particular desired goal number $g$.
For this, we use another external input to excite the \emph{goal} neuron to begin firing.
And, we input the goal number $g$ in unary, and thereby trigger the number neuron $rep(g)$ to fire.

But now we have a danger of confusion involving the number neurons.
Namely, after these initializations, both the \emph{current-number} and \emph{goal} Working Memory neurons are firing, together with their corresponding number neurons.
So there is a danger of confusion between the goal number and the current number.  We need a way to disambiguate this.
As noted in Section~\ref{sec: model: wm}, this probably involves a synchronization protocol, such as alternating the firing activities for the two Working Memory neurons.

\paragraph{Stopping:}
The mechanism must also recognize when the current count has reached the goal number $g$.  This requires an equality-detection mechanism to recognize that the  \emph{goal} and \emph{current-number} Working Memory neurons are bound to the same number neuron.  The design of such a mechanism would depend on how we are resolving the ambiguity in the bindings.
For example, if we are simply alternating the firings of the two pairs of neurons,
then the mechanism would have to recognize when the same number neuron fired at consecutive times.  This might be implemented by maintaining some firing history in the state of each number neuron, or in special neurons as in~\cite{HitronLMP20}.
%



When the count is finished, we might also want to stop the firing of the \emph{current-number} neuron.
For this, we should use some inhibitory input.  This remains to be worked out.

\vspace{.5cm}
For Query 2, starting the count is slightly different, in that an external input is used to trigger the firing of the starting letter neuron.  Starting the \emph{current-number} neuron, and ending the count, are as before.  Stopping is as before.

\subsection{Answering the queries}
\label{sec: tie-together} 

The complete solution to Query 1 involves advancing through the number and letter sequences in parallel, until the equality mechanism detects that the \emph{current-number} neuron and the \emph{goal} neuron bind to the same number neuron. 

At that point, computational activity shifts to the IKS:  another external excitatory input signal gets sent to all the intuitive concept neurons in the IKS that correspond to letter neurons in the SKS.
The unique currently-firing letter neuron in the SKS also contributes potential to its corresponding intuitive concept neuron in the IKS.
Based on a combination of these two types of potential, the correct intuitive concept neuron for the chosen letter starts firing.

Once the correct intuitive concept neuron starts firing, it causes a cascade of firings in the IKS, which yields a probability distribution of decision neurons, or emotional reaction neurons, that are reached by the cascade.  Under plausible assumptions about the IKS, with high probability, this soon leads (within some time $t_{iks}$) to a unique, persistent decision.
Analyzing this probability will require a more detailed model of how the firing cascades operate in the IKS.

The overall latency for answering the query will be approximately $g d + t_{iks}$, for some constant $d$, where we are allowing $d$ steps to account for each increment of \emph{current-number}.

Processing for Query 2 is similar, and we leave that to the reader.  The difference is in the initialization of \emph{current-number}.

\section{Parse Trees}
\label{sec: parsing}

The second example of a symbolic knowledge structure is a parse tree for a sentence.
A parse tree, as studied in linguistics, consists of nonterminals at the internal nodes of a tree and terminals at the leaves.
Parse trees are generally constructed based on context-free grammar rules. 
Since these rules are recursive, this allows for the construction of arbitrarily large and complex parse trees.  However, sentences that are normally encountered in practice are not very large or complex, so their parse trees are rather small.

In this paper, I consider a variant of the usual parse trees, in which the children of every node in the tree are unordered.  For example, a sentence may consist of a subject noun, predicate verb, and object noun, but the order of these three elements is not significant.  In different languages, these three elements might be presented, orally or in writing, in different orders---say, (subject, predicate, object) vs. (subject, object, predicate).  But in the SKS representation, it is enough to remember just that the sentence consists of these three parts.  

Unlike the sequences in Sections~\ref{sec: sequences} and~\ref{sec: diseases}, I do not assume that the parse trees are memorized in the SKS ahead of time.
Rather, the words of a sentence are presented as input (by hearing or reading), and the parse tree gets constructed on-the-fly.
However, I think it is very likely that the general structure of typical parse trees is built into the brain a priori, as a result of evolution or early learning.  For example, for 3-word sentences that consist of just a subject, predicate, and object, there is probably a built-in structure in the SKS that represents this general parse tree structure. The particular nouns and verb that appear in a particular sentence are, of course, not built in, but are filled in during processing of the sentence. 
For more complex sentences, more work will be needed during sentence processing to determine the structure of the tree, as well as for filling in the words.

In any case, during sentence processing, an (unordered) parse tree gets constructed in the SKS.  Some simplified representation of this tree then gets passed to the IKS, which associates intuitive meaning with the sentence via cascades of firing, producing as output a kind of ``story''.  Here I am drastically simplifying matters, by neglecting the possibility that the IKS helps in constructing the parse tree.  This sort of semantic disambiguation is common in language understanding, but I will restrict attention for now to sentences that can be parsed based just on their syntax.

\paragraph{Example:  3-word sentences:}
As a special case, I will sometimes consider parse trees with a very simple structure corresponding to 3-word sentences consisting of a subject noun, predicate verb, and object noun.

In terms of traditional context-free grammars, the nonterminals are Sentence, Subject, Predicate, Object, Noun, and TransitiveVerb, with Sentence as the starting nonterminal.  The set of terminals is a large collection of already-memorized nouns and transitive verbs.
The formal grammar productions are:
\begin{itemize}
    \item Sentence $\Rightarrow$ Subject Predicate Object\footnote{But recall that the order of the three elements is not significant.  We should more properly regard the right-hand side of this production as a set consisting of these three nonterminals. This simplification will probably affect the representations and their processing costs.}.
    \item Subject $\Rightarrow$ Noun, Predicate $\Rightarrow$ TransitiveVerb, Object $\Rightarrow$ Noun.
    \item Noun $\Rightarrow$ boy, baby, horse, ball, banana, tablecloth,...; Verb $\Rightarrow$ kicks, eats, sews,...
\end{itemize}
In this special case, verbs are always present tense.
Example sentences include ``Boy kicks ball.'', ``Baby eats banana.'', ``Horse sews tablecloth.'', etc. 
Assuming this general parse tree structure is built into the SKS a priori, what remains to be done during parsing is just filling in the terminals.

\vspace{.5cm}
For more general cases, I will consider more complicated sentences and parse trees, and also cases where the sentence structure is not known a priori, but must be determined during sentence processing.
In this latter case, additional work will be involved in determining the parse tree structure, in addition to the work of filling in the terminals.

Some details appear in the next two subsections.  But this is all quite preliminary, and there is much more to be done.

\subsection{Representation and computation in the SKS}
\label{sec: rep-comp-SKS}

Here I describe how parse trees might be represented in the SKS, how the parsing process might work, and what output is produced.

\paragraph{Parse tree representation:}
A parse tree could be represented in the SKS fairly directly, with a neuron representing each node of the tree and edges connecting the neurons that represent parents and children in the tree.  Note that this neural network representation does not specify any ordering for the children of a node, which is fine since we are considering unordered versions of parse trees.  The terminal and nonterminal symbols are also represented by neurons in the Lexicon; for each such symbol that occurs in a parse tree, there are edges back and forth between the Lexicon entry and the neurons in the parse tree that are labeled by that symbol.

The parse tree for an input sentence can be constructed in the SKS by allocating neurons for the nodes in the parse tree and establishing the needed connections.  This may seem like a time-consuming learning process, which is inconsistent with the speed with which humans recognize basic sentences.  However, the process can be sped up if the high-level portions of the tree are already represented in the SKS, as a result of evolution or previous learning.  This captures the idea that the brain starts the parsing process already having built-in familiarity with standard sentence structures.  I think this is reasonable, and I will assume it from now on.  

In fact, to make things as simple as possible, I will assume here that the SKS maintains a collection of standard sentence structures, each with a complete parse tree shape and all of its nonterminals filled in.  Moreover, every sentence that is input will fit one of these standard structures.  So the only things that need to be filled in are the terminals.\footnote{This is definitely an oversimplification of real sentence processing, but I think it is worth understanding this case first.}

\paragraph{The parsing process:}
Given our strong assumptions, the parsing process needs to perform only two tasks:  (1) determine which standard structure fits the given sentence, 
and (2) finish the construction of the parse tree by associating specific terminals to the leaves.

Since sentences are input word-by-word, it seems plausible that the SKS might start out with a large collection of possible sentence structures, and narrow down the set of possibilities with each new word, until only one possible structure remains.  For each new word that is input, the SKS tries to incorporate the word into its proper place in each of its still-possible parse trees.
%
If it fails (for example, because the new word is the wrong part of speech for the structure), then the SKS drops the structure from its set of possibilities.

For example, consider two sentence structures: (subject, predicate, object) and (subject, predicate). 
Suppose that an English sentence is being presented, and its second word is a verb that can be either transitive or intransitive (such as eats, runs, reads, etc.).
Then after the first two words arrive, both structures are still possible, and the verb can be assigned to a leaf in both of their parse trees.  After that, either the sentence ends,
or another noun arrives.  This leads to attempts to match both partial parse trees, but only one succeeds.


It remains to devise a workable SNN-based data structure for keeping track of the set of possibilities.  This should include: a way of representing multiple parse trees in the SKS; a mechanism to incorporate each successive input into each possible parse tree, producing an indicator when this is impossible; and a mechanism to identify when only one parse tree remains in the set of possibilities.

During construction of a parse tree, the Working Memory will help in assigning terminals to particular leaves of the tree.  For example, in a simple (Subject, Predicate, Object) sentence, a Working Memory neuron may represent the Subject role, and may bind to a symbol neuron for a particular noun.  This binding should persist long enough to establish the firing patterns needed to incorporate the noun into the SKS representation of the parse tree.\footnote{The use of Working Memory will be more complicated in more complicated sentences.
Consider the sentence ``The boy who eats the candy pets the duck.''
The phrase ``The boy who eats the candy'' can be parsed first, with the help of Working Memory neurons.  The result of this parsing is a branch of a parse tree.
But then we want to treat the entire phrase as a unit, to be related to the rest of the sentence.
So a nonterminal symbol node in the SKS that represents this entire phrase can bind to a Working Memory neuron that corresponds to the role ``Subject'', and the rest of the parse can proceed.  
In fact, the same neurons in Working Memory might be reused in different parts of the parse, e.g., a Subject role neuron can bind to a neuron representing the terminal ``boy'' in parsing the first phrase, and later to a neuron representing the subtree for the entire phrase ``The boy who eats the candy''.}

Although the internal SKS representations for parse trees are language-independent, the method of assigning terminals to nodes in the parse trees is language-specific.
For example, consider simple three-word (subject, predicate, object) sentences.
Recall that the parse trees do not specify an order for these three parts of the sentence.  But particular languages will have their own ordering conventions.
For example, in English, the first word in a sentence will generally be the Subject, the second the Predicate, and the third the Object.
Other languages may use different conventions.
These language conventions lead to different rules for parsing an input sentence.
In this way, we can have a language-independent internal representation of the sentence structure, with a language-specific way of performing the parse.

This all suggests to me that the internal SKS parse tree representations may be built-in, as a result of  evolution, whereas particular languages are certainly learned.
It remains to develop particular learning algorithms by which the brain can learn different parsing methods for different languages, all based on the same SKS parse tree structures.


\paragraph{Output of the SKS:}
In any case, the SKS eventually produces the parse tree for an input sentence.
%
%
Actually, it probably doesn't retain the entire parse tree; for example, the intermediate neurons representing the nonterminals Noun and TransitiveVerb are not necessary once parsing is finished.  
So what the SKS produces is a reduced version of the parse tree;
for example, for a simple 3-word sentence, the reduced version will probably contain some representation of the sentence as a whole, of the three parts of the sentence, and of the actual nouns and verbs that make up these parts.\footnote{
This representation might be a small labeled graph of neurons, containing a neuron for the sentence and one for each of its three terminals.  Edges between the sentence neuron and the terminal neurons could be labeled by their roles---subject, predicate, or object.}

\paragraph{Arbitrarily complex sentence structures:}
For standard sentence structures, the representations of the parse trees are compact and fit easily in the SKS.
Also, the Working Memory is large enough to keep track of the parts of the sentence and manage the parsing process.
But this is not the case for arbitrarily complex sentence structures.
%
In order to construct very large parse trees, the parts that don't fit in the SKS and Working Memory would have to be recorded somewhere outside the SKS, perhaps on paper.  There is no limit to how much can be written on paper!


\subsection{Representation and computation in the IKS}

Now what happens to the (reduced) representation of the parse tree that is produced by the SKS?  
Basically, the firing of the neurons in this representation, possibly with the help of some external signal, should trigger the firing of some related intuitive concept neurons in the IKS.  Then the IKS firing cascades proceed until they result in some output.

The intuitive concept neurons that are triggered should comprise some type of outline for the ``story'' that the parsed sentence tells.\footnote{
This may bring to mind Winston's emphasis on ``stories'' produced in the brain~\cite{Winston15}.  However, Winston's stories are different:  they are paragraphs consisting of many short sentences, from which a computer program can extract meaning.  It is possible that the representations used in this computer programs might provide ideas for how simple stories can be represented in an IKS model.}
This should be a collection of neurons that together represent the individual components and the relationships between them; the precise structure for this outline remains to be determined.
In the case of simple 3-word sentences, we might have one concept neuron that represents the general idea of a ``story'' plus neurons representing the three concepts that comprise the story, tagged somehow by their roles in the story (as in the SKS representation). But there may be more to it:  a verb might be represented, not by a single concept neuron, but by a collection of neurons that fire according to some timing pattern corresponding to the action denoted by the verb.  ``Kicks'' would probably be represented by some neurons whose firing corresponds to the action of kicking, including its direction and speed. 

\paragraph{Open question 2:} Pin down precise reduced representations for parse trees in the SKS, and precise representations for outlines for simple stories in the IKS.  Develop a mechanism by which an SKS representation of a parse tree can trigger the corresponding IKS representation of a story outline.
\vspace{.5cm}

Once these initial concept neurons are triggered, they start the usual IKS cascade of firing, continuing until some IKS output is produced.  This cascade does not follow just the associations that arise from the individual elements of the sentence/story, but also the combinations of the elements and the relationships between them.
That is, the cascade selects for associations that are relevant to the story.
For example, in the sentence ``Boy kicks ball.'', the combination of words is likely to trigger a picture of a boy aged 6-10 running outdoors, kicking a soccer-ball-sized ball, rather than a picture of a baby boy sitting indoors, or a golf ball on a tee.  The sentence ``Horse sews tablecloth.'' is likely to trigger a cartoon of a horse sitting at a table, but is unlikely to trigger a picture of a racehorse at a racetrack or an elegant lace tablecloth in a dinner setting.  
The context provided by the other parts of the story and their relationships influence the associations that arise from each particular part.  

Such a cascade pattern should arise naturally in a Spiking Neural Network model because the neurons representing the various parts of a story all contribute potential to other concept neurons.  This potential adds up, thereby encouraging the firing of neurons for concepts that are relevant to the entire story, rather than just the individual parts.

The IKS representation of the story gets refined over a short period of time, as a result of the cascade of IKS neuron firings.
The IKS representation that is triggered directly by the SKS representation, that is, the story outline, is probably rudimentary, and gets more detailed over a short time, as more context information gets incorporated.
All of this is still fast thinking, since the actions of the IKS cascade are fast.

What is the final output that is produced by this IKS process? 
It should be a mental picture of the activity in the story.
Such a picture might be represented in the IKS using a typical internal form for visual processing in computer vision---an abstract two-dimensional picture that has rudimentary representations of the key components (stick figures, outlines of buildings, etc.).  It can be a kind of cartoon.
The picture will usually be dynamic, since one of its constituents is a verb, indicating an action.

As we had for the sequence example, we could have a further IKS output representing the main judgment or emotion that is triggered by the picture, such as ``pleasant'', ``unpleasant'', or ``absurd''.

\paragraph{Interaction between SKS and IKS:} 
We have been considering only a special case of sentence understanding in which the SKS performs symbolic processing first, then passes control to the IKS, which produces a picture 
of the story, and judges whether the story is pleasant, unpleasant, absurd, etc.  
This one-way interaction is similar to what happens in our sequence example.
More elaborate parsing examples would incorporate feedback from the IKS to the SKS, for instance, when semantic information is used to resolve ambiguities in the parse structure.




\section{Learning the Representations}
\label{sec: learning}

So far, I have described situations in which symbolic knowledge structures have already been established.
But we should also consider how these structures might become established in the first place, by evolution or learning.  
Some structures may be innate, such as a neural structure that is particularly suited for learning arbitrary sequences or hierarchical structures, or for recognizing certain standard types of sentences.
But many other structures must be learned from experience.
I won't go into depth here about learning issues, but will simply give a high-level overview.

Here are some general types of things that must be learned.  In all cases, we can consider the changes at two levels of abstraction, and how they correspond at the two levels.

\begin{enumerate}
    \item  \emph{New intuitive concepts:}  
    These must get placed somewhere in the IKS.  New neuron(s) must be selected to represent them.  This might involve the use of a Winner-Take-All mechanism.  Valiant's book~\cite{Valiant} contains some example protocols based on random graphs.  Lynch and Mallmann-Trenn~\cite{LynchMallmannTrenn21} present others based on identifying the candidate neuron with the highest incoming potential.
    
    \item  \emph{New symbols:}  
    These must get placed somewhere in the SKS, and also in the Lexicon.  The method of choosing new representing neurons can be similar to what is used in the IKS.  Also, for each new symbol, connections must be set up in both directions between the neurons in the Lexicon and those in the SKS corresponding to the new symbol. 
    
    \item  \emph{Connections between intuitive concepts:}  
    Since people (and probably other animals) can learn to relate any pair of intuitive concepts, the brain must have sufficient physical connections between concept neurons to support this learning. 
    These connections can be strengthened or weakened as a result of experiences.
    
    Thus, the underlying directed graph in the IKS should also be sufficiently connected.
    Edge weights should increase and decrease in response to experiences, according to some version of Hebbian learning.  This learning will probably be gradual because of the noisy nature of learning of intuitive concepts and relationships.

    \item  \emph{Connections between symbols:} 
    Learning a symbolic structure, such as a sequence or linguistic structure, is different from learning connections between intuitive concepts.
    It doesn't just happen automatically, as a result of experience, but requires attention, is more purposeful, and is controlled by some outside stimulus.  
    The process should be less noisy than intuitive learning.
    
    Thus, learning in the SKS should also involve attention, in the form of Working Memory.
    The learning rule does not need to accommodate much noise, so it can make larger jumps than learning rules in the IKS.
    
    Because activity in the SKS is more purposeful, learning can proceed in more interesting ways than just simple Hebbian rules.
    For example, learning a sequence, such as the Greek alphabet, 
    is not just simple Hebbian learning of individual connections between consecutive pairs of symbols.
    Rather, it may be more like learning a song, taking advantage of some overall structure, such as a rhythmic pattern.
    
    For another example, consider the problem of learning a language, specifically, the sub-problem of learning the language's conventions for the order of words in a sentence, as described in Section~\ref{sec: rep-comp-SKS}.
    In terms of an SNN model, this means learning a network that will handle sentence inputs correctly for the given language, assigning the successive words to the appropriate positions in a built-in parse tree. 
    This network might be learned through a supervised learning process, where wrong attempts lead to nonsense sentences and give negative feedback.


    \item
    \emph{Connections between symbols and their intuitive counterparts:}  
    Intuitive concepts may be learned simultaneously with their symbols, or in either order.  In any case, a new symbol is learned (in the SKS and Lexicon), and a new concept is learned (in the IKS).  Repeated simultaneous presentation of the symbol and the concept can lead, via Hebbian learning, to strengthening of weights on the edges connecting them.
    %
    Another matter that remains to be considered is how connections can be learned between complex symbolic structures and corresponding complex intuitive structures. 
    

\end{enumerate}

In all of these cases, it would be good to describe and analyze specific learning mechanisms.  These could be evaluated in terms of speed of learning.

\section{Conclusions}

In this paper, I have proposed that two distinct types of structures are present in the brain:  \emph{Symbolic Knowledge Structures (SKSs)}, used for formal symbolic reasoning, and \emph{Intuitive Knowledge Structures (IKSs)}, used for drawing informal associations.  I have described many ideas for modeling these structures, and for implementing them in Spiking Neural Networks.  I have developed two basic examples of their use:  counting through a memorized sequence of disease variants, and understanding simple sentences with predetermined structure. 

I repeat my disclaimer from the beginning of the paper, that this is a collection of ideas for a theory, but not a complete, coherent theory.
I am interested in developing this further, until the whole thing makes sense in terms of a formal model like that in~\cite{LynchMusco-arxiv21}.
The approach should be algorithmic, like that used in~\cite{LynchMP-arxiv19,LynchMallmannTrenn21},
with formal network descriptions and analysis.
But that will take considerably more thought and work.

Much more remains to be done in pinning down the various individual mechanisms, and expressing them in terms of a single Spiking Neural Network model.  
In particular, for the Working Memory attention mechanism, we need more details for how the binding between Working Memory neurons and SKS neurons works, especially for how firing is synchronized (Open question 1).
For the parsing example, we need more details of how a parse tree gets filled in, and how it gets copied from the SKS to the IKS in the form of a story outline (Open question 2).

Many more examples of symbolic reasoning can be considered, including more interesting arithmetic calculations than just counting, and more elaborate linguistic examples than parsing of sentences with simple, predictable structure. 
Higher mathematical reasoning such as solving algebraic equations, and even creative construction of mathematical proofs, are also interesting to consider.
We can also study examples involving elaborate interactions between SKS and IKS, rather than just simple one-way interactions as in the examples in this paper.

So, there is still very much to be done.
I hope that this preliminary paper might elicit some suggestions, and interest some other researchers in pursuing these ideas.

\section{Acknowledgments}
I thank Sabrina Drammis and Brabeeba Wang for listening to my preliminary ideas and offering  various helpful suggestions.

\bibliographystyle{plain}
\bibliography{SKS-IKS-biblio}

\end{document}